\documentclass[fleqn,twoside]{article}
\usepackage{espcrc2}

\usepackage{graphicx}
\usepackage[figuresright]{rotating}

\newcommand{\AmS}{{\protect\the\textfont2
  A\kern-.1667em\lower.5ex\hbox{M}\kern-.125emS}}

\title{~~~~~Early results of the LHCf Experiment and their contribution to\\
~~~~~~~~~~~~~~~~~ Ultra-High-Energy Cosmic Ray Physics}

\author{~O.~Adriani(1,2),~L.~Bonechi(1),M.~Bongi(1),G.~Castellini(1,3),R.~D'Alessandro(1,2),~A.~Faus(4),~K.~Fukatsu(5),\\~M.~Haguenauer(6),~Y.~Itow(5),~K.~Kasahara(7),~D.~Macina(8),~T.~Mase(5),~K.~Masuda(5)\\
~Y.~Matsubara(5),~H.~Menjo(1),~G.~Mitsuka(5),~Y.~Muraki(9)\thanks{This paper has been presented at CRIS2010 by Y. Muraki.~Correspondence email address:~muraki@stelab.nagoya-u.ac.jp, muraki@konan-u.ac.jp },~M.~Nakai(7),~K.~Noda(5),~P.~Papini(1),\\~A-L.~Perrot(8),~S.~Ricciarini(1),
~T.~Sako(5),~K.~Suzuki(5),~T.~Suzuki(7),~Y.~Shimizu(7),\\~K.~Taki(5),
~T.~Tamura(10),~S.~Torii(7),~A.~Tricomi(11,12), ~J.~Velasco(4),
~W.~C.~Turner(13),~K.~Yoshida(14)\\
~~~~~~~~~~~~~~~~~~~~~~~~~~~~~~~~~~~~~~~~~~~~ ---~~ The LHCf collaboration~~ --- 
\\
1)~INFN Sezione di Firenze, Italy\\
2)~Universit\`a degli Studi di Firenze, Florence, Italy\\
3)~IFAC CNR, Florence, Italy\\
4)~IFIC, Universitat de Val\`encia, Valencia, Spain\\
5)~Solar-Terrestrial Environment Laboratory, Nagoya University, Nagoya, Japan\\
6)~Ecole-Polytechnique, Palaiseau Cedex, France\\
7)~RISE, Waseda University, Tokyo, Japan\\
8)~CERN, Gen\`eve, Switzerland\\
9)~Department of Physics, Konan University, Kobe, Japan\\
10)~Institute of Physics, Kanagawa University, Yokohama, Japan\\
11)~Universit\`a degli Studi di Catania, Catania, Italy\\
12)~INFN Sezione di Catania, Catania, Italy\\
13)~Accelerator and Fusion Research Division, LBNL, Berkeley, USA\\
14)~Faculty of System Engineering, Shibaura Institute of Technology, Saitama, Japan}

\begin{document}

\begin{abstract}
LHCf is an experiment dedicated to the measurement of neutral particles emitted in the
very forward region of LHC collisions.  The physics goal is to provide data for calibrating
hadron interaction models that are used in the study of Extremely High-Energy Cosmic-Rays.
The LHCf  experiment acquired data from April to July 2010 during commissioning time of LHC 
operations at low luminosity.  Production spectra of photons and neutrons 
emitted in the very forward region ( $\eta>$ 8.4) have been obtained.
In this paper preliminary results of the photon spectra taken at  $\sqrt{ s}$ = 7TeV are reported.
\vspace{1pc}
\end{abstract}

\vspace{5mm}
\maketitle

\begin{sloppypar}

\section{Introduction}

Since it is impossible to directly measure the energy of ultra-high-energy 
cosmic rays ($\sim$ $10^{20}$ eV), the cosmic ray primary energy is 
traditionally determined by measurements of the secondary nuclear and
electromagnetic cascade showers that are produced in the atmosphere 
by the primary cosmic rays. Comparison of the shower measurements 
with MC simulations then gives the primary energy. In order for this to work
reliably it is necessary to know the very forward neutral particle 
(most importantly neutral pion) production cross-section in the energy range 
of the primary cosmic rays. Since theoretical models currently predict 
a rather large range of cross-section values, one must look to accelerator data at the 
highest possible energy in order to get reliable information for calibrating the MC codes.

   The world's highest energy particle accelerator for at least the next few decades 
is the LHC at CERN which began operation in 2009 and has so far operated at center
of momentum proton-proton collision energy $\sqrt{s }$ = 0.9 and 7 TeV.  It is currently 
projected that in 2013 the LHC will produce proton-proton collisions at its maximum 
energy $\sqrt{s }$= 14 TeV, which is equivalent to $10^{17}$eV for a cosmic ray proton 
entering the atmosphere. The LHCf experiment [1] has been designed 
to provide accelerator data for calibration of the MC codes that are used 
for the simulation of ultra-high-energy cosmic ray showers. Specifically 
LHCf will measure the production spectra of very forward neutral pions 
and neutrons in the rapidity range $\eta$ $>$ 8.4. Some details of the experimental 
arrangement are given in Section 2 below.

   To explain the cosmic ray experiment and MC situation in somewhat more detail, 
in a typical air shower experiment the total number of shower electrons and 
positrons (Ne) is measured as a function of atmospheric depth X in gm/$cm^{2}$. 
The number of shower particles at the shower maximum, $N_{e,max}$ is then multiplied 
by $\sim$2 GeV to give an estimate of the primary energy. Of course the factor $\sim$2 GeV 
is approximate and depends on details of the altitude of observation, the incident 
angle of the cosmic ray and fluctuations in the starting point of nuclear interactions.  
All these details are treated by the MC codes. 

   In Figure 1 we show some experimental cosmic ray data and MC simulations [2]. 
The vertical axis is the depth of the shower maximum in gm/$cm^{2}$ and the horizontal 
axis is the cosmic ray primary energy as it enters the atmosphere.  From Figure 1 
we see that the DPMJET2 model predicts that the composition of cosmic rays at the 
super high-energy region (say $>$ $10^{19}$eV), is dominated by heavy primaries such as 
iron nuclei. On the other hand, the QGSJET01 model predicts that cosmic rays in 
this region are dominated by protons.  This discrepancy can be resolved by calibrating 
the simulation codes with accelerator data at the highest energy and is the motivation 
for the LHCf experiment.      

   In high energy hadronic collisions the energy flow is dominated by the very 
forward emitted particles and shower development is dominated by the production 
of electromagnetically interacting particles (photons, electrons and positrons). 
The most important cross section for cosmic ray shower development is therefore 
the forward production of neutral pions in hadron collisions which then immediately 
decay to two forward photons. Until now the highest energy of an accelerator experiment 
measuring forward production of neutral pions was attained by the UA7 experiment [3] 
at the CERN pbar-p collider, the laboratory equivalent energy of UA7 was 1.5$\times$ $10^{14}$eV, 
far below the $10^{19}$eV region of interest for present day cosmic ray physics.  
As noted above, when LHC reaches its design energy of 14 TeV in 2013 the LHCf experiment  
will extend the energy range for which there is accelerator data by three orders of 
magnitude to $10^{17}$eV. Although this is still two to three orders of magnitude below 
the highest energy cosmic rays, the discrepancies in MC calculations are evident by $10^{17}$eV. 
In addition to helping resolve the ultra-high-energy cosmic ray composition question the LHCf 
data will be useful for resolving questions at the GZK cut-off region around $10^{19}$eV. 

  Parenthetically we note that in Figure 1 all the MC calculations are in reasonably 
good agreement at $10^{14}$eV corresponding to the energy of the UA7 experiment where 
accelerator calibration data is available. The MC results diverge the higher one 
gets above $10^{14}$eV. The goal of LHCf is to reduce the uncertainty of the MC calculations 
at $10^{17}$eV to something resembling the current situation at $10^{14}$eV.

  In large collider experiments, the detectors (ATLAS, CMS etc.) are traditionally 
designed for detection of new massive particles in the central high $p_{T}$ region.  
Vacuum beam pipes in the very forward region make it difficult to put particle 
detectors there. However the shape of the beam pipe in the LHC collider makes 
it relatively easy to insert neutral particle detectors at zero degree collision angles.  
At $\pm$140m from the IP the beam pipe makes a Y transition from a single common beam pipe 
facing the IP to two separate beam pipes joining to the arcs of LHC. 
The TAN neutral particle absorbers are located at the crotch of the Y and contain 
a 90 mm wide instrumentation slot between the two beam pipes where the LHCf detectors 
are inserted.  
    In the remainder of this paper we will describe some details of the LHCf detectors 
and the very early results that have been obtained.
  
\begin{figure}[htb]
\vspace{9pt}
\includegraphics[width=7.5cm]{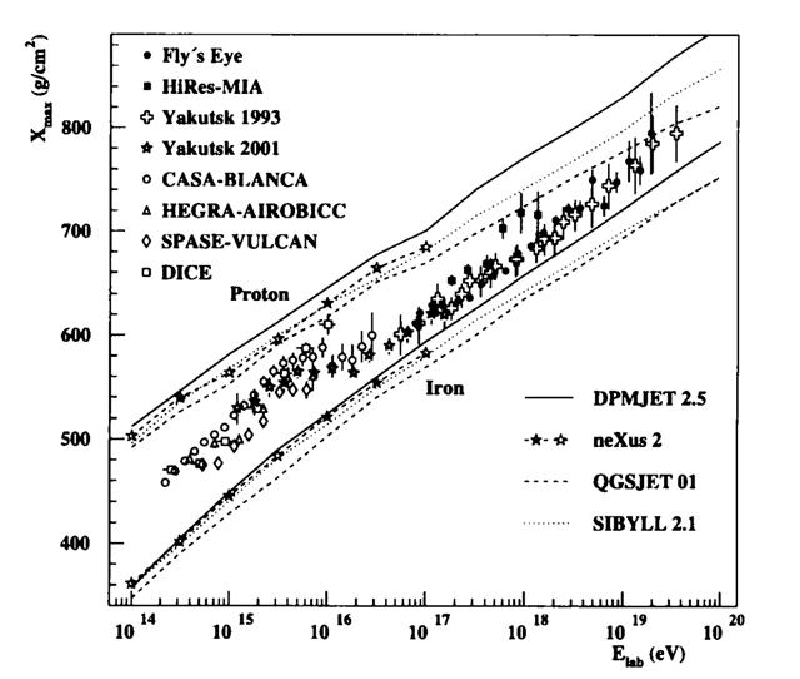}
\caption{The shower maximum in gm/$cm^{2}$ versus the primary cosmic ray energy 
in eV for various experiments and MC simulations [2].}
\label{Fig1.eps}
\end{figure}

\section{The detectors}

   The LHCf detector consists of two detectors, Arm 1 and Arm 2 located on opposite sides 
of the collision point IP1. Each detector consists of two calorimeters.  
The calorimeters consist of 44 radiation lengths (1.7 strong interaction lengths) 
of tungsten plates interleaved with sixteen 3 mm thick layers of plastic scintillator 
for sampling the shower intensity. In addition there are four X-Y layers of position 
sensitive detectors for measuring the transverse location of the shower axis. 
Arm 1 uses SciFi for the position sensitive layers whereas Arm 2 uses micro 
strip silicon sensors. The position resolutions have been measured 
with an electron beam at the SPS, obtaining 160$\mu$m for the Arm 1 SciFi and 49$\mu$m 
the Arm 2 micro strips.

\begin{figure}[htb]
\vspace{9pt}
\includegraphics[width=7.0cm]{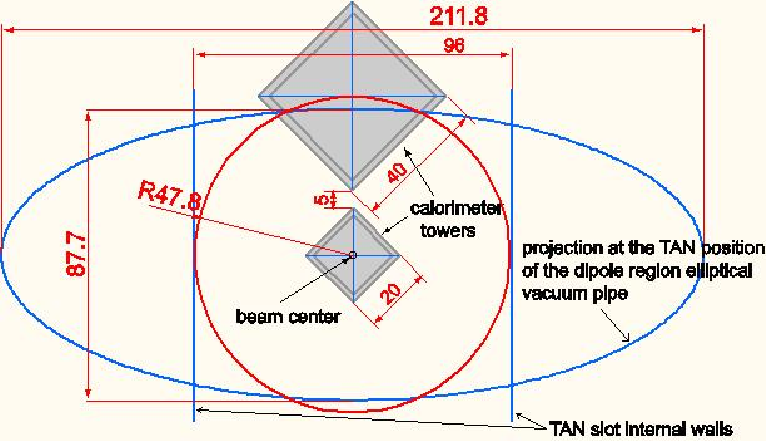}
\caption{Transverse cross sections of the LHCf detector for Arm 1.}
\label{Fig2.eps}
\end{figure}

\begin{figure}[htb]
\vspace{9pt}
\includegraphics[width=7.0cm]{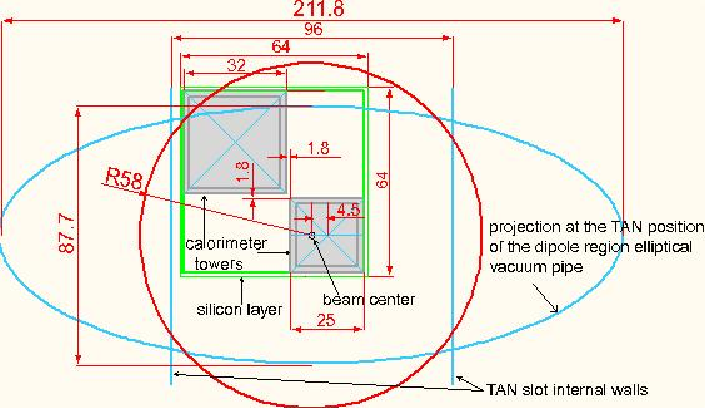}
\caption{Transverse cross sections of the LHCf detector for Arm 2.}
\label{Fig3.eps}
\end{figure}

   As shown in Figures 2 and 3, the Arm 1 detector consists of 2cm$\times$2cm and 4cm$\times$4cm 
cross section calorimeters while the Arm 2 consists of 2.5cm$\times$2.5cm 
and 3.2cm$\times$3.2 cm calorimeters.  The calorimeters are installed in the 
instrumentation slots of the TAN absorbers $\pm$140m from IP1 and at zero degree 
collision angle. The calorimeters together with light pipes and PMTs are contained 
in aluminum boxes that can be remotely moved up and down relative to the beam height. 
In addition to the calorimeters shown in Figs. 2 and 3, each Arm has a thin plastic 
scintillator in front of it for estimating relative luminosity.
 
   The calorimeter signals are sent to the underground USA15 counting area via 180 m cables. 
The DAC trigger is generated in the USA15 area and recorded events are tagged 
according to whether or not an event was also recorded by ATLAS. Frequent calibrations 
of the gain of the scintillator and the photomultipliers have been carried out with 
laser light pulses sent from the USA15 area. Compared to the initial installation 
in the autumn of 2009, a 3$\%$ degradation of scintillator sensitivity was observed 
by the end of the run on July 19, 2010.
 
   After the LHCf detectors were first installed in the autumn of 2009, 
p-p collision data were first taken with $\sqrt{s}$=900 GeV.  
Collisions at energy $\sqrt{s}$=7000 GeV were started in April, 2010.  
LHCf data has been taken at low luminosity and during machine commissioning 
in order to limit "pile-up events" (multiple collision events during the same bunch crossing).  
The typical pile up rate was as $\leq$0.7$\%$ in the middle of June under 
the luminosity of L=5.5 $\times$$10^{28}$/$cm^{2}$sec for the 3$\times$3 bunch beam crossing scheme.  
The beam-gas collision rate was estimated during the $\sqrt{s}$=900 GeV run to be 0.1$\%$.

\section{The energy calibration of calorimeters}

In order to reduce the number of events with two or more neutral particles 
striking a single calorimeter the LHCf detectors have been designed 
with the small cross sections shown in Figures 2 and 3. A consequence of these small 
cross sections is that there is some shower leakage out of the edges the calorimeters. 
The Moliere radius of tungsten is about 1 cm. As a consequence, for a shower centered 
on the 2cm$\times$2cm calorimeter, approximately 16$\%$ of the shower particles 
at the shower maximum leak outside the edges of the calorimeter. This leakage 
effect has been thoroughly investigated with electron beams at the SPS. 
The electron beam data have been used to create tables of shower leakage correction 
factors as a function of the position of the shower axis. In order to limit 
the magnitude of the correction factor below 7$\%$, events 
having a shower axis within 2 mm of the calorimeter edges are eliminated 
from the analysis. The details of the shower leakage effect and correction factors 
have been published in a previous report [4].
The absolute energy scales of the calorimeters have been calibrated 
with SPS test beams and MC simulation. The MC code EPICS v.8 predicts 
that a 100 GeV electron deposits 430 MeV in the fourth layer of scintillator 
which is at the shower maximum. The corresponding flux of minimum 
ionizing particles (MIPs) is $\sim$940. On the other hand the experimentally 
measured ADC for this layer was 295.   This means that one MIP corresponds 
to about 295/940 = 0.314 ADC channel counts or 0.00785 pC. 
The photomultiplier (PMT) voltage was 450 V for the electron beam data. 
For muon beam data, which corresponds to a single minimum ionizing particle, 
the PMT voltage was increased to 1000 V corresponding to a PMT 
gain increase of $\sim$100, The observed ADC was 30 whereas the MC predicted 35. 

    The linearity of the PMT plus scintillator combinations have been checked 
with a laser light beam over a dynamic range that corresponds to fluxes 
from 1 MIP to 2$\times$ $10^{5}$MIPs. Excellent linearity was obtained by proper choice 
of PMT (Hamamatsu-R7400U) and a long decay constant scintillator (EJ-260). 
We have also used actual beam data to confirm the linearity of the conversion 
tables between the ADC and the number of minimum ionizing shower particles.  
We have not observed any saturation effects on the outputs of the PMTs. 
The technical details of the linearity studies are given in reference [4].  
The final calibration of the calorimeter energy scales will be made 
by using the invariant mass peaks of neutral pion and eta mesons.

 The energy resolutions of the calorimeters have been investigated 
using SPS electron beams from 50 to 200 GeV.  The MC that replicated the 
SPS results has been used to estimate the energy resolution at 1 TeV, 
obtaining 2.8$\%$. Further details of the energy resolution 
can be found in reference [5].

\section{Data analysis and some early results}

    We would first like to demonstrate that the Arm 1 and Arm 2 detectors 
are working similarly. For this demonstration only photons that hit the detectors 
within the same rapidity range have been selected. 
Figure 4 shows the photon spectra obtained by the Arm 1 (red points) and Arm 2 
(blue points) detectors. Photons emitted within a radius 5 mm from the beam axis are plotted. 
It is evident that the photon spectrum produced 
in proton-proton collisions with $\sqrt{s}$=7000 GeV can be measured 
very well by either detector independently.  Minor differences
are noted. These may be due to alignment corrections for the centers of the 
two proton beams relative to the detectors that have not yet been applied. 

\begin{figure}[htb]
\vspace{9pt}
\includegraphics[width=7.5cm]{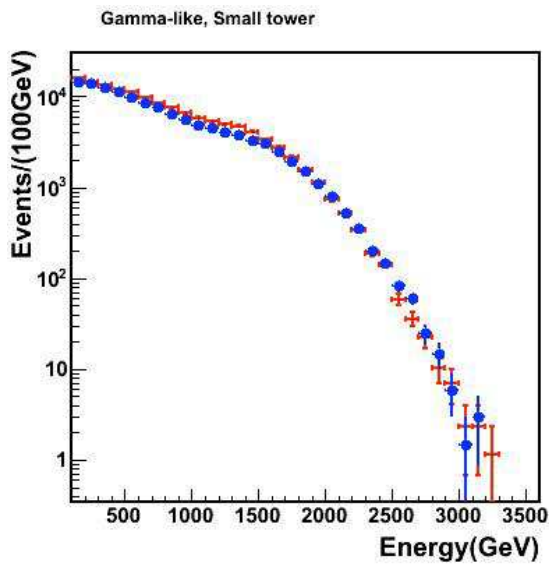}
\caption{The photon spectrum observed by the small towers 
of Arm1 (red +) and Arm2 (blue $\bullet$).}
\label{Fig4.eps}
\end{figure}

    Both photons and neutrons enter the LHCf detectors and produce showers. 
Separating the photons and neutrons is an important task for LHCf. We have 
developed a separation criterion based on the fact that, owing to the order 
of magnitude difference in radiation and strong interaction lengths, on average 
neutron showers initiate and penetrate much deeper into the calorimeters than 
photon showers. However at intermediate depths there is some overlap between 
the photon and neutron showers. In calorimetry this particle separation 
problem usually occurs as e/p separation but in our case it is $\gamma$/n separation. 

   The $\gamma$/n particle identification (PID) method has been developed with MC simulations 
based on the QGSJET2 and DPMJET3 models.  In Figure 5, we present the results 
of MC simulation for photon and neutron induced showers together with experimental 
data from LHCf.  The parameter L90$\%$, in units of radiation length, represents the depth 
in which 90$\%$ of the shower energy has been deposited and is shown on the horizontal axis.  
The vertical axis is the number of events having a particular value of L90$\%$.  
From Fig 5 we see that events with L90$\%$ $<$18 are essentially all photons, 
events with L90$\%$ $>$ 22 are essentially all neutrons and events 
with 18 $<$ L90$\%$ $<$ 22 
are a mixture of photons and neutrons. 

  Selecting events according to a particular value of L90$\%$ 
invariably involves a weak energy dependence.  
We have made MC simulations for three cases of PID criteria for selecting 
photon showers; (case 1) we fix the length of L90$\%$ at 18 r.l. or less, 
(case 2) we fix the detection efficiency of photons to be greater than 90$\%$, 
and (case 3) the photon detection efficiency is fixed at 95$\%$ and the length L90$\%$ 
is increased to 30 r.l..  Case 3 is most conservative in terms of not eliminating 
photon showers but has the highest neutron contamination. However for all cases, 
if appropriate corrections are applied to the raw data, the MC simulations predict 
that we can infer the contamination free photon spectra at $\sqrt{s}$ =7000 GeV.
This is shown in Figure 6.

    In the last part of this section we would like to describe 
in more detail the corrections due to shower leakage. The case 
when only a single photon is incident on the Arm1 or Arm2 detectors 
has been described above. The case when two photons are incident, 
one in each of the two calorimeters of Arm1 or Arm2 is a little more involved. 
In that case the photon energies $E_{\gamma1}$ and $E_{\gamma2}$ are related to 
the energies $E_{1}$ and $E_{2}$ measured by the calorimeters by\\
~~$E_{\gamma1}$~=~$E_{1}$~+~$\Delta$$E_{1}$~-~$\Delta$$E_{2,1}$\\
~~$E_{\gamma2}$~=~$E_{2}$~+~$\Delta$$E_{2}$~-~$\Delta$$E_{1,2}$\\
~~where the subscript 1 refers to the small calorimeter and subscript 2 
to the large calorimeter of a given detector. $\Delta$$E_{1}$ and $\Delta$$E_{2}$ are 
the corrections due to shower leakage out of a given calorimeter described previously.  
$\Delta$$E_{2,1}$ is the shower leakage out of calorimeter 2 that leaks into calorimeter 1 
and similarly for $\Delta$$E_{1,2}$.  These correction factors have been estimated 
by the MC calculations.  To limit the magnitude of the corrections events 
are rejected from analysis if a shower axis falls within 2 mm of the edge of a calorimeter. 
The fractions of shower leakage out of a given calorimeters that leak into 
the other calorimeter are typically $\sim$20$\%$.  So, for example if there is 16$\%$ 
leakage of photon 1 shower energy out of calorimeter 1, there is about 3$\%$ leakage 
of photon 1 shower energy into calorimeter 2.

   The two photon correction process that has been described will have 
an effect on the measured mass of the neutral pion that is dependent 
on the pion energy. The reason for the pion energy dependence is that 
the mean opening angle between the photons decreases as the pion energy increases.   
As the opening angle decreases the shower axes will tend to fall closer 
to the calorimeter edges and therefore the magnitudes of the leakage corrections 
will increase. The magnitude of the effect has been estimated with MC simulations. 
For proton-proton collisions with $\sqrt{s}$=7 TeV the leakage corrections increase 
the measured neutral pion mass by 2.5$\%$ at the lowest pion energy ($\sim$900 GeV) 
and 5$\%$ at the highest pion energy ( $\sim$1.3TeV) detected by the Arm1 and Arm2 detectors.  

   At the time of this report our data analysis does not yet 
separate out multi-hit events in a single calorimeter and pile-up 
events with more than a single detected proton-proton interaction 
in a bunch crossing. For these reasons it is too early to present 
our measurements together with MC predictions. 

\begin{figure}[htb]
\includegraphics[width=8.0cm]{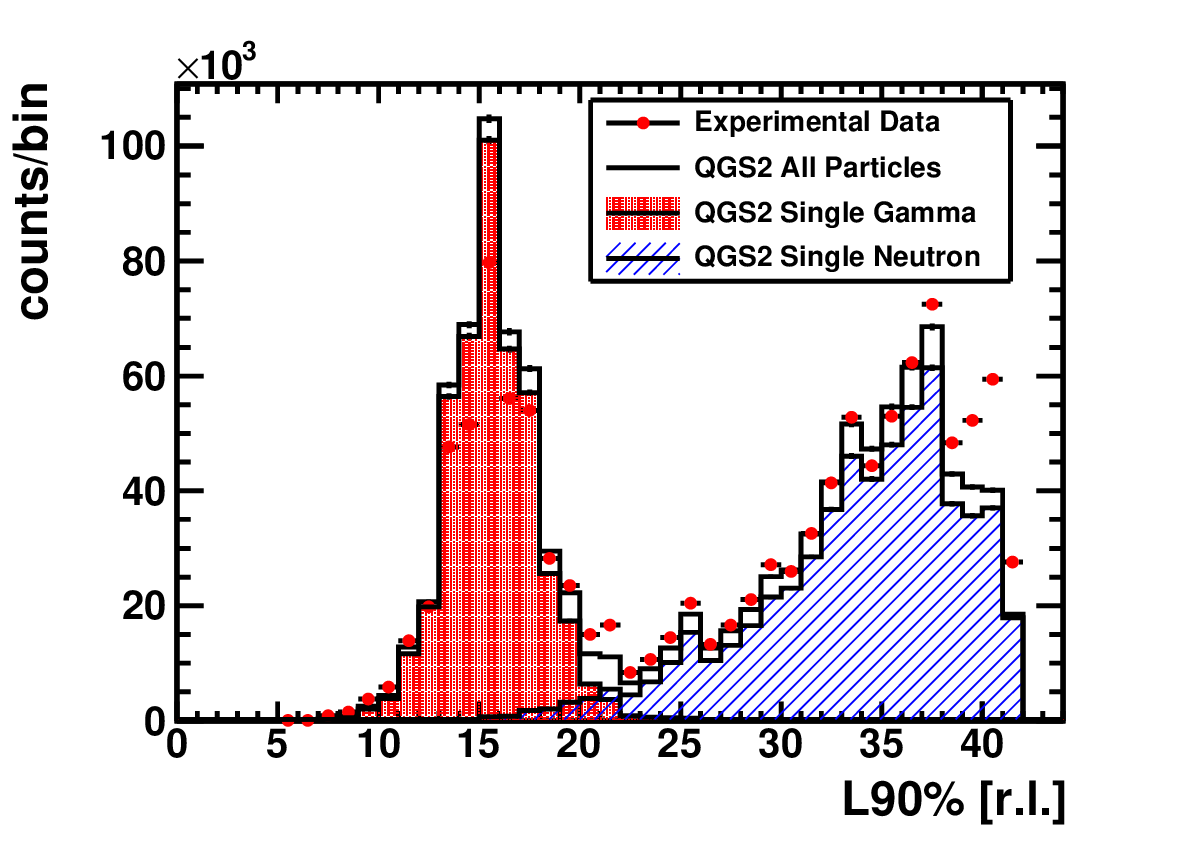}
\caption{Particle Identification between photons and neutrons.
The MC results are shown together with the real data. 
L90$\%$ means that 90$\%$ of shower particles are involved in
the radiation length expressed in the horizontal value.  }
\label{Fig5.eps}
\end{figure}

\begin{figure}[htb]
\includegraphics[width=8.0cm]{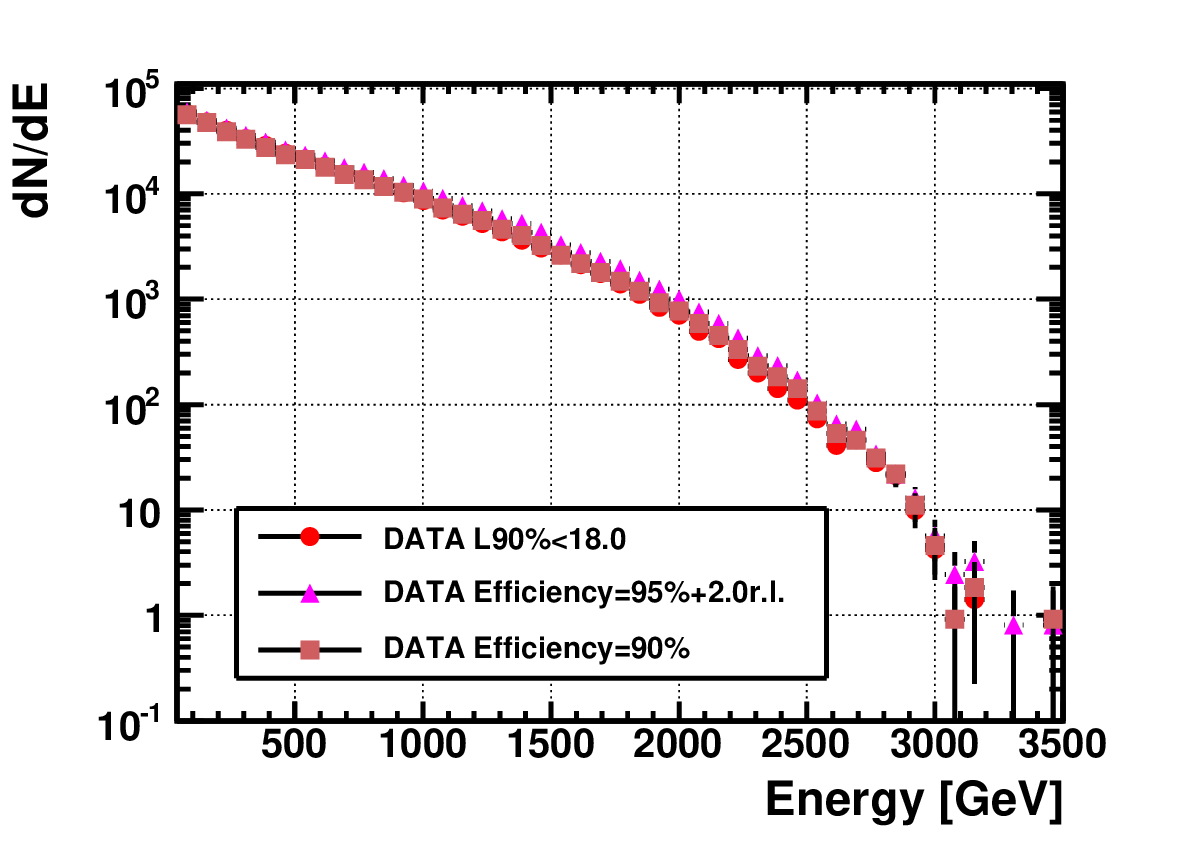}
\caption{Photon spectra are plotted using three different PID criteria
for separating photons from neutrons.   The three criteria give essentially
the same result and indicate insensitivity to which of the criteria is used. }
\label{Fig6.eps}
\end{figure}

\section{Summary of the experiment}

(1) The LHCf experiment has collected high energy single photon ($\sim$200 million events 
for each arm) and neutral pion ($\sim$one million events for each arm) events during 
low intensity operation of the LHC.  These numbers of events are sufficient for 
completing LHCf data analysis at 7 TeV operation of LHC so the LHCf detectors 
have been removed until the energy of LHC is increased in the future. 

(2) The LHCf detectors have been calibrated with SPS beams, by laser light 
and by utilizing the invariant mass peak of the neutral pion.

(3) A particle identification procedure (PID) for separating photons and 
neutrons has been developed.  Calorimeter leakage corrections have been 
thoroughly investigated with SPS electron beam measurements and MC simulations.
 
(4) Data analysis is in a preliminary stage however it may be stated 
that the shape of the photon production cross-section is well described 
by the SIBYLL and EPOS codes, while the neutron production cross-section 
is well described by the QGSJET and DPMJET models.  Discussion of the relative 
merits of the models that are customarily used by cosmic ray physicists 
for simulation of super high energy events must wait until     
we obtain differential cross-sections.\\

 \section{Acknowledgments}  
The authors acknowledge Drs.~Karsten~Niebuhr, 
Michelangelo Mangano  Prof. Mario Calvetti for 
useful discussions in the early stage of the project.

\end{sloppypar}

\end{document}